*batteries* Article

# Optimizing Energy Management and Sizing of Photovoltaic Batteries for a Household in Granada, Spain: A Novel Approach Considering Time Resolution

Catalina Rus-Casas *,1,2, Carlos Gilabert-Torres [1,2] and Juan Ignacio Fernández-Carrasco [1,2]

1 Electronic Engineering and Automatic Department, University of Jaén Las Lagunillas Campus, A3 Building, 23071 Jaén, Spain; gilabert@ujaen.es (C.G.-T.); jifernan@ujaen.es (J.I.F.-C.)
2 Centre for Advanced Studies in Energy and Environment CEACTEMA, Universidad de Jaén, 23071 Jaén, Spain
* Correspondence: crus@ujaen.es
**Abstract:** As residential adoption of renewable energy sources increases, optimizing rooftop photovoltaic systems (RTPVs) with Battery Energy Storage Systems (BESSs) is key for enhancing self-sufficiency and reducing dependence on the grid. This study introduces a novel methodology for sizing Home Energy Management Systems (HEMS), with the objective of minimizing the cost of imported energy while accounting for battery degradation. The battery model integrated nonlinear degradation effects and was evaluated in a real case study, considering different temporal data resolutions and various energy management strategies. For BESS capacities ranging from 1 to 5 kWh, the economic analysis demonstrated cost-effectiveness, with a Net Present Value (NPV) ranging from 54.53 € to 181.40 € and discounted payback periods (DPBs) between 6 and 10 years. The proposed HEMS extended battery lifespan by 22.47% and improved profitability by 21.29% compared to the current HEMS when applied to a 10 kWh BESS. Sensitivity analysis indicated that using a 5 min resolution could reduce NPV by up to 184.68% and increase DPB by up to 43.12% compared to a 60 min resolution for batteries between 1 and 5 kWh. This underscores the critical impact of temporal resolution on BESS sizing and highlights the need to balance accuracy with computational efficiency.

**Citation:** Rus-Casas, C.; Gilabert-Torres, C.; Fernández-Carrasco, J.I. Optimizing Energy Management and Sizing of Photovoltaic Batteries for a Household in Granada, Spain: A Novel Approach Considering Time Resolution. *Batteries* 2024, *10*, x. https://doi.org/10.3390/xxxxx

Academic Editors: Rodolfo Dufo-López and Pascal Venet

Received: 19 July 2024
Revised: 13 September 2024
Accepted: 7 October 2024
Published: date
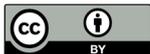

**Copyright:** © 2024 by the authors. Submitted for possible open access publication under the terms and conditions of the Creative Commons Attribution (CC BY) license (https://creativecommons.org/licenses/by/4.0/).
## 1. Introduction

The electricity demand in the residential sector increased by 17% between 2000 and 2020, now accounting for 29.9% of the total electricity consumed in the European Union. In response, the European Union has set objectives to reduce its $CO_2$ emissions by 40%, compared to 1990 levels, and to increase the share of renewable energy consumed to 32% by 2030 [1].

To achieve these goals, the use of renewable energies has been intensified with plans such as REPowerEU. This plan highlights the crucial role of solar PV, which is expected to reach an installed capacity of 320 GW by 2025 and 600 GW by 2030 [2]. The contribution of RTPVs to this installed capacity is significant, reaching nearly 50% in countries such as Spain. RTPV systems promote the implementation of distributed generation (DG) within the EU's electricity systems [3]. This increase in photovoltaic generation has been fostered by a significant reduction in the cost of these installations, with a global value reduction of 52.2% from 2016 to 2024 [4].

*Batteries* **2024**, *10*, x. https://doi.org/10.3390/xxxxx                    www.mdpi.com/journal/batteries

However, photovoltaic systems face technical challenges, primarily due to their intermittent nature. Variability in solar power generation can affect power system stability and power quality, leading to power flow fluctuations [5,6].

In addition to technical challenges, RTPVs face a significant economic challenge due to the depreciation of surplus photovoltaic energy, caused by the large-scale injection of PV into the grid. In many regions, solar power generated during peak irradiation hours exceeds local demand. This has led to an unprecedented decrease in electricity prices during these periods, negatively affecting feed-in tariffs. This depreciation of the photovoltaic surplus, which reached a value of 60% between 2023 and 2024 in Spain [7], reduces the profitability of solar installations. Consequently, this negatively affects the economic viability of photovoltaic projects and discourages new investments in this type of clean energy.

To address these problems, BESSs have been proposed as a promising solution with great potential. Several studies show that lithium iron phosphate batteries offer greater fiscal benefits compared to other types of batteries [8,9].

There are several topologies of battery installations, each adapted to different applications and energetic needs. Among the most common are utility-scale installations. These high-capacity storage systems are located in front of the meter (FTM). The main function of these batteries is to provide frequency regulation and energy backup services to the utility grid [10,11]. On the other hand, there are behind-the-meter (BTM) battery systems, which are found in residential, commercial, and industrial facilities. In these settings, batteries can be used to reduce peak demand, manage energy consumption, and improve power quality [12]. In the BTM scenario, there are homes in which the batteries store the solar energy generated during the day for use at night, thereby increasing self-sufficiency and reducing dependence on the grid [13,14].

The increase in the implementation of these battery configurations has been facilitated by the decrease in their price in recent years [15]; however, they remain relatively expensive. In most cases, the inclusion of batteries in the design of an installation requires detailed technical and economic study. Studies in the industrial sector demonstrate the profitability of using batteries for peak shaving, where peaks occurring during regular operations result in significant penalties on the electricity bill [16,17]. However, in the residential sector, the studies carried out do not always confirm the economic viability of these installations [18]. Other studies in the residential field propose the use of BESS energy management strategies in energy communities to analyze their feasibility [19].

Among the most commonly employed strategies by the HEMS to optimize battery usage is the maximization of self-consumption. In this case, the aim is to increase local self-consumption and reduce energy imported from the grid. The BESS stores surplus solar PV energy during periods when demand is lower than PV generation and delivers it during periods when demand is higher than PV generation [20].

Another concept that appears in battery energy management is energy arbitrage. This concept refers to the management of energy consumption and supply in the context of fluctuating prices [21]. In a real-time pricing (RTP) tariff, as seen in a day-ahead market, prices vary throughout the day, reflecting the surplus or deficit of power generation and other factors, such as grid congestion. These variations can be significant, and a BESS provides the flexibility to consume energy during off-peak hours and deliver it to the household during peak hours. This results in financial savings for the consumer and reduces grid congestion during peak hours. For consumers on time-of-use (TOU) tariffs, where there may be two or more fixed prices during the day, the battery can be programmed to charge during low-price periods and reduce consumption during high-price periods. Control in this case is simpler than with an RTP tariff, but the economic benefits are lower [22,23].

The literature presents a variety of optimization methods for applying different energy management strategies within the HEMS. These methods can be broadly classified into heuristic methods, exact methods, and mixed methods [24]. Among the heuristic methods are applications of the Genetic Algorithm [25,26] and Particle Swarm Optimiza-tion [27,28]. Exact methods such as Linear Programming [29,30], Non-Linear Programming [31,32], or Dynamic Programming [33,34] are also applied to obtain the optimal energy dispatch. Multi-objective mixed methods are used for the optimization of multiple objectives simultaneously, such as self-consumption, profitability, and avoided $CO_2$ emissions [35,36]. The most common objective in the literature is to minimize the cost of energy, aiming to maximize the profitability of the investment [24].

The scientific studies reviewed highlight various aspects to consider in RTPVs that incorporate batteries, focusing on both economic aspects and the selection of optimization strategies for managing energy flows. These strategies have been approached with varying levels of detail and do not always address all the specificities of battery use. An important consideration when sizing and evaluating the long-term economic profitability of these investments is the aging mechanisms of the batteries [37].

Therefore, it is essential to account for battery degradation due to cycling and time, which can be influenced by factors such as time, temperature, and the operating profile of the BESS [38,39]. A semi-empirical model for evaluating battery degradation, based on theoretical analysis and experimental studies, has been developed in [40]. While previous works have incorporated degradation costs into energy dispatch, most do not account for nonlinearities in battery degradation or, if they do, update the battery capacity every several months or even yearly [41,42]. This leads to a significant overestimation of battery capacity and, consequently, its flexibility in optimizing energy dispatch, distorting the results.

It is a technological challenge to know the real degradation of the battery and it can go unnoticed if the HEMS data has time resolutions that do not allow knowledge of the real degradation. Table 1 presents studies regarding the optimization of BESS sizing.

**Table 1.** Main findings of the literature on BESS sizing for dwellings. Source: self-elaborated.

| Ref. | Sampling Period | Scheduling Optimization Objective | Optimization Time Resolution | Optimization Period |
|---|---|---|---|---|
| [23] | 60 min | Min. Operation Cost | 60 min (average daily profile by season and weather conditions) | 1 day |
| [31] | 60 min | Min. Total Cost | 60 min (average daily profile) | 365 days |
| [41] | 60 min | N/A | N/A | N/A |
| [30] | 60 min | Min. Energy Cost | 60 min | 365 days |
| [43] | 60 min | Min. Energy Cost and Investment Cost | 60 min | 365 days |
| [18] | 15 min | N/A | N/A | N/A |
| [42] | 15 min | Max. Savings and min. degradation cost | 15 min | 0.5 day |
| [44] | 1 min | Min. Operation Cost | 1 min | 365 days |
| This work | 5 min | Min. Energy Cost and Degradation cost | 5 min | 7 days |

As shown in Table 1, the time resolution typically ranges between 15 and 60 min, and the impact of this parameter on final results is not assessed. When evaluating battery degradation, the time resolution of energy dispatch can lead to a loss of information that affects the estimation of battery degradation and, consequently, its profitability. Although some studies use higher data resolutions, they often do not account for battery aging when optimizing energy management [44].

The novelty of this study lies in the development of a methodology based on an energy management optimization algorithm, implemented using MILP, and aimed at deter-mining the optimal sizing of BESSs for households with pre-existing photovoltaic installations. This methodology is applied to a real case study to identify the optimal BESS size, enabling an evaluation of the current BESS size and a detailed assessment of the advantages and disadvantages of the proposed HEMS. Additionally, the case study was monitored at a temporal resolution of 5 min, and battery degradation was calculated on a weekly basis, facilitating a precise techno-economic analysis of different BESS sizes. This study focuses on lithium batteries, specifically LFP technology, which is the most commonly used in the RTPVs field. The objective of this article is to propose a new method for the management and sizing of lithium BESSs and to evaluate the influence of time resolution and HEMS operation on techno-economic results. The analysis assesses various techno-economic scenarios using real data from an actual installation: a residential RTPV connected to the grid and a DC-coupled battery. To minimize energy dispatch costs, both theoretical and empirical models for predicting and assigning costs to battery degradation were reviewed. An optimization model based on mixed-integer linear programming (MILP), widely used in the scientific literature, was employed [30,43]. Finally, this method was applied across different time resolutions, considering commercial battery models compatible with the pre-installed inverter.

The results from the case study indicate that battery capacities ranging from 1 to 5 kWh are cost-effective, and that energy management via the proposed HEMS enhances the profitability and extends the lifespan of the BESSs in comparison with the existing HEMS. Specifically, for the optimal battery size (2 kWh), an NPV of €181.40 and a payback period of 7.4 years were obtained. For economically viable BESSs, the sensitivity analysis of temporal resolution revealed an overestimation of lifespan and NPV of up to 20.61% and 184.68%, respectively, highlighting its importance as a critical parameter when performing techno-economic evaluations of these systems.

## 2. Materials and Methods

### 2.1. Description of the System

The installation under study is a grid-connected single-family house located in the province of Granada, in southern Andalusia, Spain. It has a PV generator and a LFP battery coupled in DC to the inverter. Figure 1 illustrates a schematic of the HEMS.

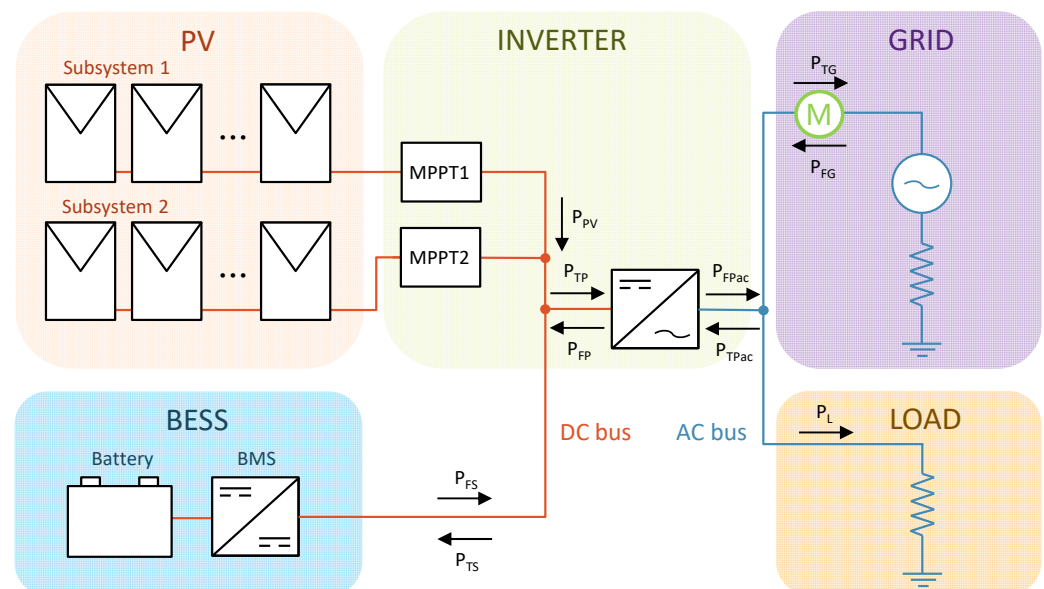

**Figure 1.** Schematic diagram of the system. Source: self-elaborated.

The main characteristics of the equipment comprising the system are detailed in Table 2. The photovoltaic generator is composed of two subsystems, each using the same PV module, with a total peak power of 7.4 kW.

**Table 2.** Characteristics of the system equipment.

| Equipment | Specification | Value |
|---|---|---|
| PV modules | Technology | Polycrystalline silicon |
| | Max. rated power (STC [1]) | 275 W |
| | Number | 27 |
| | Efficiency | 16.90% |
| Inverter | Max. PV input power | 9 kW |
| | Number of MPPTs | 2 |
| | Rated output power | 6 kW |
| | European efficiency | 97.8% |
| BESS | Chemistry | LFP |
| | Effective battery capacity | 10 kWh |
| | Max. charge/discharge power | 5 kW |
| | Max. depth of discharge (DOD) | 100% |
| | Roundtrip efficiency | 94% |

[1] STC according to Standard IEC 61724-1 [45].

The installed inverter is capable of delivering a maximum AC power of 6 kW and is equipped with two maximum power point trackers (MPPTs) that allow for the connection of the two photovoltaic subsystems. The inverter also features two terminals for the DC coupling of the battery. The house is equipped with a commercial lithium–ion BESS that includes a 5 kW power control module and two storage modules, each with a capacity of 5 kWh.

*2.2. Monitored Data*

For this study, data from the year 2023 was utilized, encompassing a full year of operation. To ensure accurate characterization of the PV generation and consumption of the dwelling, the data sampling period was set to 5 min. Variables such as DC and AC PV generation, along with the exchanged power of the battery and its state of charge (SOC), were derived from data recorded by the inverter using its commercial software. Additionally, the system was equipped with a high-precision wattmeter (Class 1) to measure the energy exchange between the house and the grid, as shown in Figure 1. This setup allowed for precise measurement of the remaining energy flows in the system, allowing the energy demand of the home to be calculated according to Equation (1).

$$P_L = P_{PV,ac} + P_G \tag{1}$$

AC PV generation and home consumption are shown in Figure 2, aggregated by hour and month. Regarding PV generation, it is notably higher in the spring and summer months (April–September) compared to the fall and winter months (October–March), due to greater solar radiation and longer daylight hours.

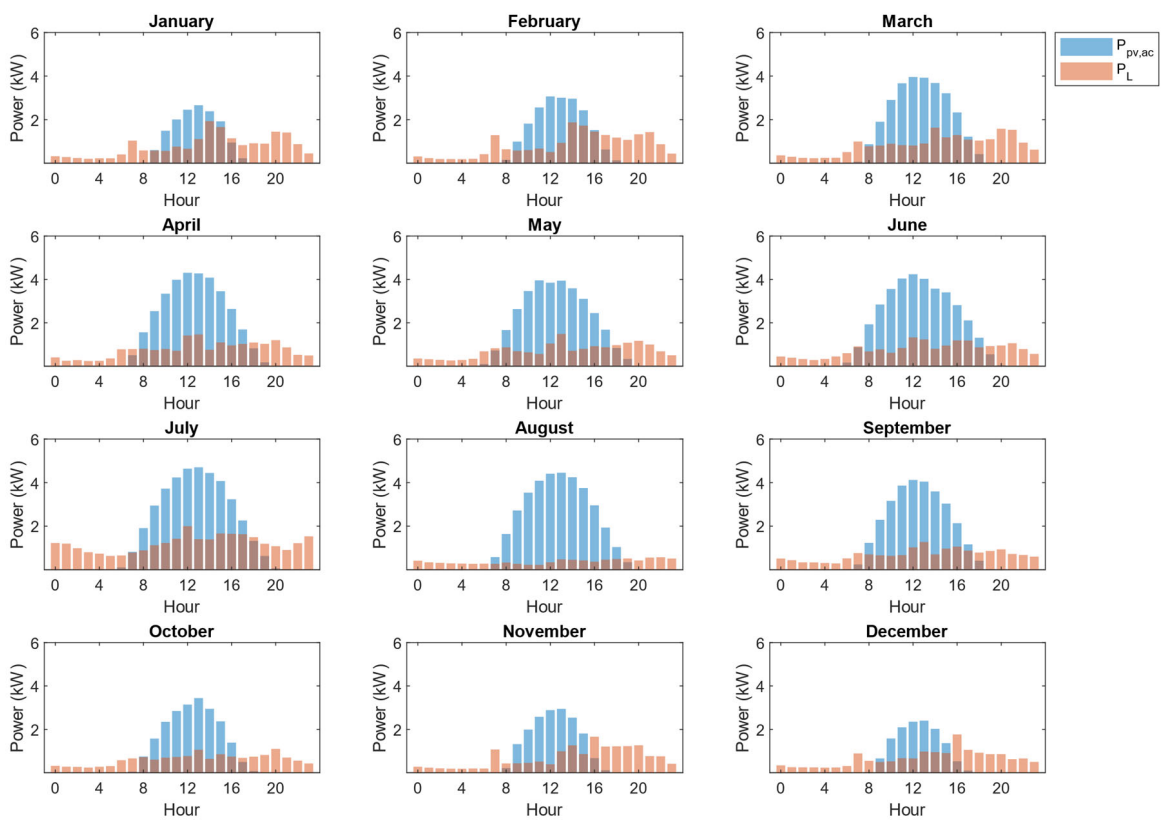

**Figure 2.** Average values of PV generation and demand per hour and month of the real case study located in Granada, Spain. Source: self-elaborated.

With regard to the consumption of the installation, the typical profile of a residential home can be observed: low consumption at night, with two peaks, during midday and evening, coinciding with the times when residents are at home. Nighttime consumption is generally low, except in July, which is attributed to the warm weather in the region during summer. Moreover, there is a significant decrease in residential consumption during the holiday period in August.

In addition to analyzing PV generation and consumption separately, it is interesting to analyze the coupling of both profiles. The analysis of this coupling, which corresponds to the portion of PV consumed, shows that the house has a self-consumption of 35.69% in the autumn and winter months, while self-consumption drops to 27.68% during the spring and summer months, when more energy is fed into the grid due to higher generation. Self-sufficiency during the year has an average value of 46.59%, reaching a maximum (57.88%) in April and a minimum (31.38%) in December.

Houses with similar locations and consumption profiles, studied in [46], achieved maximum efficiency with a generation-to-consumption ratio close to 1, which is known as the net zero energy building (ZEB) point. In our case study, the annual PV generation of the RTPV was 9.58 MWh/year, while the energy demand was 6.23 MWh/year, resulting in a generation-to-consumption ratio of 1.54. Thus, the RTPV was oversized. The inclusion of an LFP BESS enabled the utilization of solar surplus, enhancing both the self-consumption and self-sufficiency of the house. This work represents a significant advancement in updating RTPVs; if it is determined that these systems are oversized, the technical and economic feasibility of coupling a BESS in DC with a pre-installed RTPV can be evaluated to achieve greater energy self-sufficiency for the house.

*2.3. Optimization Method*

The proposed method for determining the optimal battery size for a residential installation is illustrated in Figure 3. To achieve this, a set of commercial batteries compatible with the installed inverter were considered, and the optimal power dispatch of the HEMS were simulated using MILP.

The modeling of the BESS, as in the literature of Table 1, has been performed with power and energy models (PEM), which provide a straightforward yet sufficiently accurate approach for sizing. The operating study horizon is divided into optimization periods to which the MILP is applied with the objective of minimizing the cost of energy imported from the grid, and the cost due to battery degradation. The optimization period refers to the timeframe that the optimization algorithm considers in each execution: a longer period improves the optimality of dispatch energy but requires more memory and computation time. To balance computational cost with energy dispatch optimization, a weekly optimization period (7 days) was chosen.

Energy dispatch accounts for battery degradation, estimated using the semi-empirical model developed in [40], and cycles are counted with the rain-flow counting (RFC) algorithm [47]. The simulation concludes when the battery capacity reaches a threshold value, in this case, 80% of its original capacity.

The main objective of BESS sizing is to maximize return on investment. After determining the optimal energy flows of the system, the profitability of the investment was evaluated according to the NPV and DPB parameters for each battery model. Ultimately, the model that yielded the highest profitability was selected.

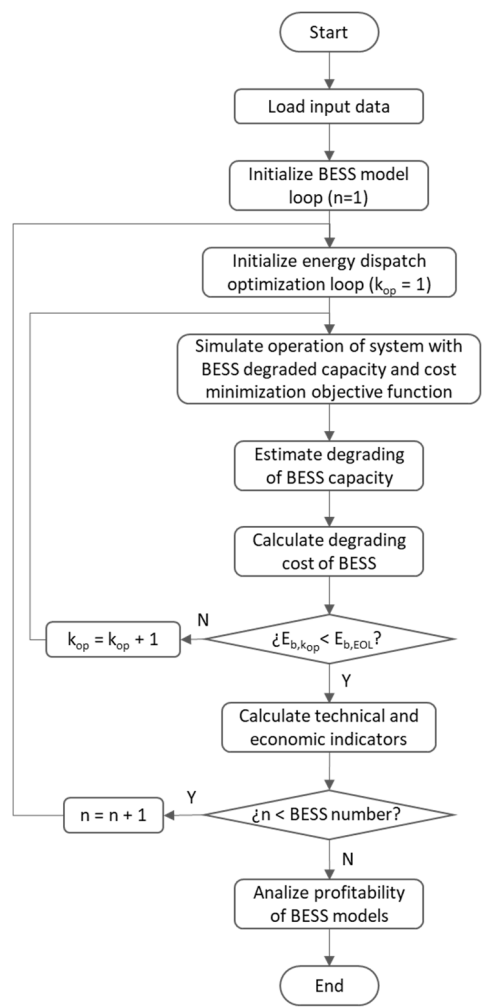

**Figure 3.** Flowchart of the proposed optimization method. Source: self-elaborated.

2.3.1. Energy Balances

The energy balance of the system in the DC bus is specified in (2), where the input and output power on the DC side of the inverter ($P_{TP,k}$ and $P_{FP,k}$) is equal to the sum of the PV generation ($P_{pv,k}$) and the power charged/discharged by the battery ($P_{TS,k}$ and $P_{FS,k}$).

$$P_{FP,k} + P_{TP,k} = P_{pv,k} + P_{FS,k} + P_{TS,k} \qquad (2)$$

In the AC bus, an energy balance is imposed (3) making the power exchanged on the AC side of the inverter ($P_{FPac,k} + P_{TPac,k}$) plus the power imported/exported to the grid ($P_{FG,k}$ and $P_{TG,k}$) equal to the power demanded by the dwelling ($P_{L,k}$).

$$P_{FPac,k} + P_{TPac,k} + P_{FG,k} + P_{TG,k} = P_{L,k} \qquad (3)$$

Inverter conversion losses were considered in the energy balance between the DC and AC buses. In Equations (4) and (5), the AC power of the inverter is calculated as a function of the DC power and the efficiency of the inverter ($\eta_{inv}$), which is considered constant.

$$P_{FPac,k} = P_{TP,k}\ \eta_{inv} \qquad (4)$$

$$P_{TPac,k} = P_{FP,k}/\eta_{inv} \qquad (5)$$

The SOC of the BESS depends on the power exchanged with the DC bus and the internal charging and discharging losses. In Equation (6), the roundtrip efficiency ($\eta_b$) is considered to account for this effect, where $t_s$ is the timestep and $E_{b,k_{op}}$ is the capacity of the BESS.

$$SOC_k = SOC_{k-1} + \frac{P_{FS,k}\ t_s}{E_{b,k_{op}}\sqrt{\eta_b}} - \frac{P_{TS,k}\ \sqrt{\eta_b}\ t_s}{E_{b,k_{op}}} \qquad (6)$$

2.3.2. Restrictions

The power on the DC bus at the inverter input is limited by maximum battery charging power ($P_{inv,dc,min}$) and maximum DC input power ($P_{inv,dc,max}$). These constraints are accompanied by the binary variable $i_P$, which indicates the direction of power flow in (7) and (8).

$$P_{inv,dc,min}(1 - i_P) \leq P_{FP,k} \leq 0 \qquad (7)$$

$$0 \leq P_{TP,k} \leq P_{inv,dc,max}\ i_P \qquad (8)$$

Analogous to the DC side of the inverter, its AC power is limited by the maximum power it is able to deliver ($P_{inv,ac,max}$). The binary variable $i_P$ indicates the direction of the power flow, as in the DC bus of the inverter, in (9) and (10).

$$0 \leq P_{FPac,k} \leq P_{inv,ac,max}\ i_P \qquad (9)$$

$$-P_{inv,ac,max}(1 - i_P) \leq P_{TPac,k} \leq 0 \qquad (10)$$

Regarding the power exchanged with the grid, its restrictions are to be less than the contracted power of the dwelling ($P_{FG,max}$) and greater than the limitation of discharge to the grid if it exists ($P_{TG,max}$). The binary variable $i_G$ indicates whether the energy flow enters or leaves the grid in (11) and (12).

$$0 \leq P_{FG,k} \leq P_{FG,max}\ i_G \qquad (11)$$

$$P_{TG,max}(1 - i_G) \leq P_{TG,k} \leq 0 \qquad (12)$$

As for the battery, its power is limited by its maximum charge ($P_{TS,max}$) and discharge ($P_{FS,max}$). The state of charge or discharge is defined by the unit variable $i_S$ in (13) and (14).

$$0 \leq P_{FS,k} \leq P_{FS,max} \; i_S \tag{13}$$

$$P_{TS,max}(1 - i_S) \leq P_{TS,k} \leq 0 \tag{14}$$

The maximum DOD of the battery is 100%, but to ensure longer life [48,49], the minimum and maximum state of charge limits were set to 0.2 and 0.8, respectively, as shown in Equation (15), resulting in a maximum DOD of 60%.

$$SOC_{min,k_{op}} \leq SOC_k \leq SOC_{max,k_{op}} \;,\; \begin{cases} SOC_{min,k_{op}} = 0.2 \; E_{b,k_{op}} \\ SOC_{max,k_{op}} = 0.8 \; E_{b,k_{op}} \end{cases} \tag{15}$$

2.3.3. Objective Function

The objective function for energy dispatch aims to minimize both the cost of energy imported from the grid and the cost of battery degradation (16). The regulated tariff VPSC or Voluntary Price for Small Customer has been chosen for this study, under which the cost of energy purchased from the grid ($C_{G,k}$) is variable (RTP) and is obtained by adding the cost of generation (hourly), charges and tolls (based on usage periods), and taxes (17). While many studies account for profit from injecting energy into the grid, this price has significantly decreased in recent years due to the surplus of photovoltaic generation during daylight hours. Therefore, income from this source was considered negligible.

The battery degradation cost $C_{bd,k_{op}}$ is computed iteratively in each optimization period and used in subsequent periods. It is calculated as the ratio between the cost of the depleted battery life and the energy discharged up to that point (18). The fraction of consumed battery life was calculated according to Equation (19), assuming an initial state of health (SOH) of 100% and an $SOH_{EOL}$ of 80%. For the initial value of $C_{bd,k_{op}}$, the battery cost was divided by the output energy guaranteed by the manufacturer.

$$\min \left\{ \sum_{k \in k_{op}} P_{FG,k} \; t_s \; C_{G,k} + \sum_{k \in k_{op}} P_{FS,k} \; t_s \; C_{bd,k_{op}} \right\} \tag{16}$$

$$C_{G,k} = GC_k + TC_k + VAT \tag{17}$$

$$C_{bd,k_{op}} = \frac{f_{bd,k_{op}-1} C_b}{\sum_{\substack{k \in r_{op} \\ k \notin k_{op}}} P_{FS,k} \; t_s} \tag{18}$$

$$f_{bd,k_{op}} = \frac{L_{k_{op}}}{1 - SOH_{EOL}} \tag{19}$$

The degraded battery capacity was calculated at the end of each optimization period according to Equation (20), where $E_{b,0}$ is the original battery capacity, $E_{b,k_{op}}$ is the degraded battery capacity and $L_{k_{op}}$ represents the capacity degradation. To determine the BESS capacity loss, the model proposed in [40] was applied, which is described in Appendix A, including the equations and parameters used.

$$E_{b,k_{op}} = (1 - L_{k_{op}}) E_{b,0} \tag{20}$$

After simulating the useful life of each battery model, the sizing criteria were evaluated, taking into account technical and economic aspects. In this way, the battery size that provides the greatest benefits can be determined.

2.3.4. Technical and Economic Sizing Criteria

Once the energy dispatch and cost data are available for each of the n batteries at their end-of-life ($T_{EOL}$), it is necessary to determine the return on investment and select the most suitable BESS model from those available. This study considered both technical and economic criteria to evaluate how battery size impacts the source of consumed energy and its profitability. From a technical perspective, the self-consumption rate (SCR) and self-sufficiency rate (SSR) were used. SCR, calculated according to Equation (21), measures the utilization of photovoltaic generation that is either directly consumed ($P_{pv-L}$) or stored in the battery ($P_{pv-bat}$) [46]. On the other hand, the SSR quantifies the direct contribution of renewable energy relative to household demand and is defined by Equation (22).

$$SCR_n = \frac{\sum_{k \in T_{EOL,n}} P_{pv-L_{n,k}} \; t_s + P_{pv-bat_{n,k}} \; t_s}{\sum_{k \in T_{EOL,n}} P_{pv_{n,k}} \; t_s} \tag{21}$$

$$SSR_n = \frac{\sum_{k \in T_{EOL,n}} P_{pv-L_{n,k}} \; t_s + P_{pv-bat_{n,k}} \; t_s}{\sum_{k \in T_{EOL,n}} P_{L_{n,k}} \; t_s} \tag{22}$$

The economic viability of investment in BESS can be assessed by indicators found in the literature such as the net present value (NPV), internal rate of return (IRR), discounted payback (DPB), and the levelized costs of electricity (LCOE) [19]. The LCOE is widely used to quantify the cost of the energy delivered by the HEMS, while the NPV quantifies the profits as the difference between the present value of the benefits and the investment cost. The breakeven point, when the NPV is zero, can be characterized by the time required in years to reach it (DPB) or the interest rate that makes the NPV of all project cash flows equal to zero (IRR). These parameters are calculated until the end of the battery's lifetime.

In this work, the economic evaluation of the investment was achieved through two parameters widely used in the literature for this purpose: NPV and DPB. Equation (23) defines the NPV of battery model $n$, where $C_{b,n}$ is the capital invested in the battery and $DCF_{n,k}$ represents the discounted cash flows.

$$NPV_n = -C_{b,n} + \sum_{k \in T_{EOL,n}} DCF_{n,k} \tag{23}$$

$$DCF_{n,k} = \frac{P^{PV}_{FG,k} \; t_s \; C_{G,k} - P^{PVB}_{FG,k} \; t_s \; C_{G,k}}{(1+i)^{y_k}} \tag{24}$$

To calculate the cost of each battery, the values reported in NREL's Annual Technology Baseline [50] were used: 252.37 €/kWh for capacity and 503.30 €/ kW for power. The discounted cash flows were calculated according to Equation (24), representing the savings between the cost of energy taken from the grid without a BESS ($P^{PV}_{FG,k}$) and with a BESS ($P^{PVB}_{FG,k}$). Here, $y_k$ denotes the number of years elapsed until step $k$, and $i$ is the discount rate, set at 5.58% in Spain for the period 2020–2025 [51]. Conversely, the DPB indicates the number of years required for the investment to become profitable, which occurs when the NPV is equal to zero.

*2.4. Case Studies*

The objective of this research is to propose a new method for optimizing battery size using MILP and to assess the impact of optimization time resolution on the cost-effectiveness of these systems in residential homes with pre-installed RTPVs. A broad range of battery sizes was considered, spanning from 1 to 21.7 kWh. While some proposed sizes did not correspond to commercially available models, actual values were used where applicable. This includes commercial battery sizes of 5, 6.9, 10, 13.8, 15, and 21.7 kWh, which were compatible with the inverter installation. Table 3 presents their characteristics:

Table 3. Energy and power capacities of models considered for the BESS sizing. Source: self-elaborated.

| BESS Model | 1 | 2 | 3 | 4 | 5 | 6 | 7 | 8 | 9 | 10 |
|---|---|---|---|---|---|---|---|---|---|---|
| Energy Capacity (kWh) | 1 | 2 | 3 | 4 | 5 | 6.9 | 10 | 13.8 | 15 | 21.7 |
| Power Capacity (kW) | 0.5 | 1 | 1.5 | 2 | 2.5 | 3.5 | 5 | 7 | 5 | 10.5 |

In order to achieve the aims of this study, the following cases were considered:

1. BESS sizing with the proposed energy management method**:** the operation of the BESS was simulated over its lifetime with a time resolution of 5 min.
2. Sensitivity analysis of temporal resolution: the lifetimes of the different BESSs were simulated with the proposed energy management method for the temporal resolutions of 5, 15, 30 and 60 min.
3. Comparison of the real operation with the proposed method: a study of the real operation of the installed BESS was conducted in order to compare and evaluate the proposed method.

## 3. Results

This section analyzes the influence of time resolution and battery power management methods on sizing BESS capacity. For this purpose, a techno-economic study was carried out for each of the cases and BESSs described in Section 2.4.

### 3.1. BESS Sizing with the Proposed Method

Regarding technical analysis of battery sizing, the mean SCR and SSR values, along with the lifespan, were obtained for the different battery sizes, considering a time resolution of 5 min. The results are presented in the scatter plot in Figure 4, along with the trend lines for each parameter.

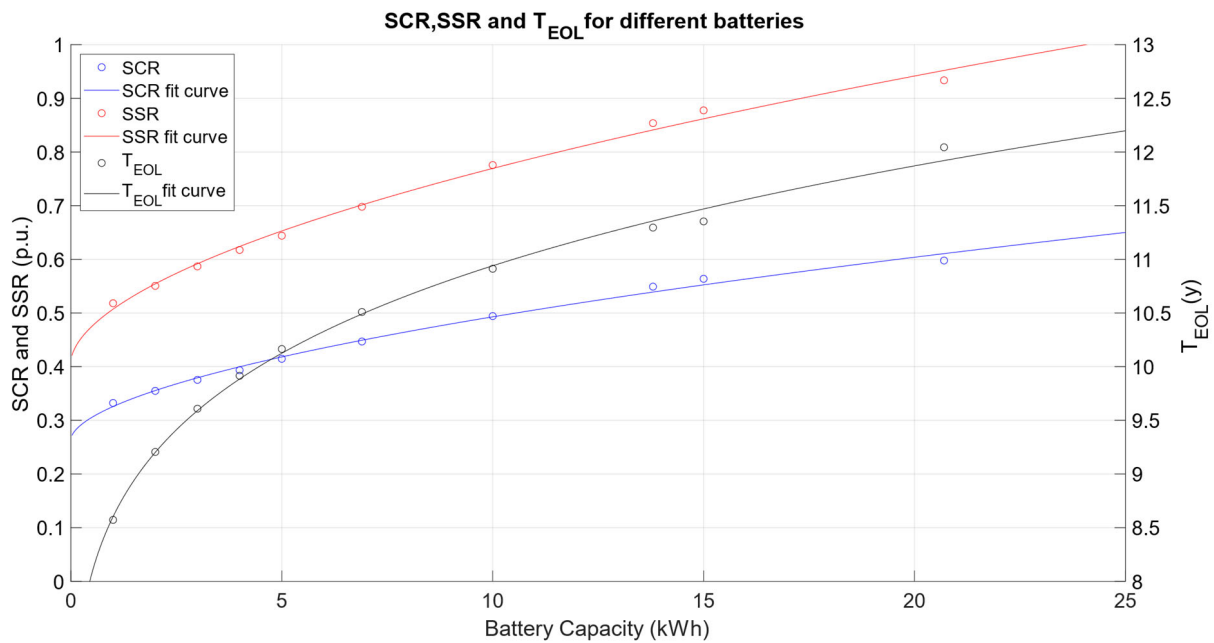

**Figure 4.** Technical analysis of battery sizing: SCR, SSR, and $T_{EOL}$. Source: self-elaborated.

Regarding battery capacity, the observed trends in SCR, SSR and $T_{EOL}$ can be divided into two sections: one where the slope is steeper (on the left) and one where the slope levels off more gradually (on the right). For both SCR and SSR, this indicates that increasing battery size is more effective when the batteries are smaller than 5 kWh, leading to greater improvements in self-consumption and self-sufficiency. However, for batteries larger than 10 kWh, the incremental benefit of increasing capacity diminishes, suggesting that the additional battery capacity is not fully utilized.

The lifetime of the BESSs ranged from 8.5 years to 12 years, increasing with greater battery capacity. Similar to self-consumption and self-sufficiency, the increase in lifespan relative to capacity is more pronounced for batteries smaller than 5 kWh. To understand why lifetime varies significantly with battery size, the RFC results were analyzed as illustrated in Figure 5. This graph displays the number of cycles performed as a function of their DOD and average SOC for three different BESS sizes: 1 kWh (Figure 5a), 5 kWh (Figure 5b), and 15 kWh (Figure 5c). For high DODs (0.6), the number of cycles is quite similar across all three battery sizes. However, for lower DODs, the number of cycles varies with battery capacity: the 5 kWh and 15 kWh batteries exhibit more cycles at a DOD of less than 5%, whereas the 1 kWh battery performs more cycles at a DOD between 5% and 10%.

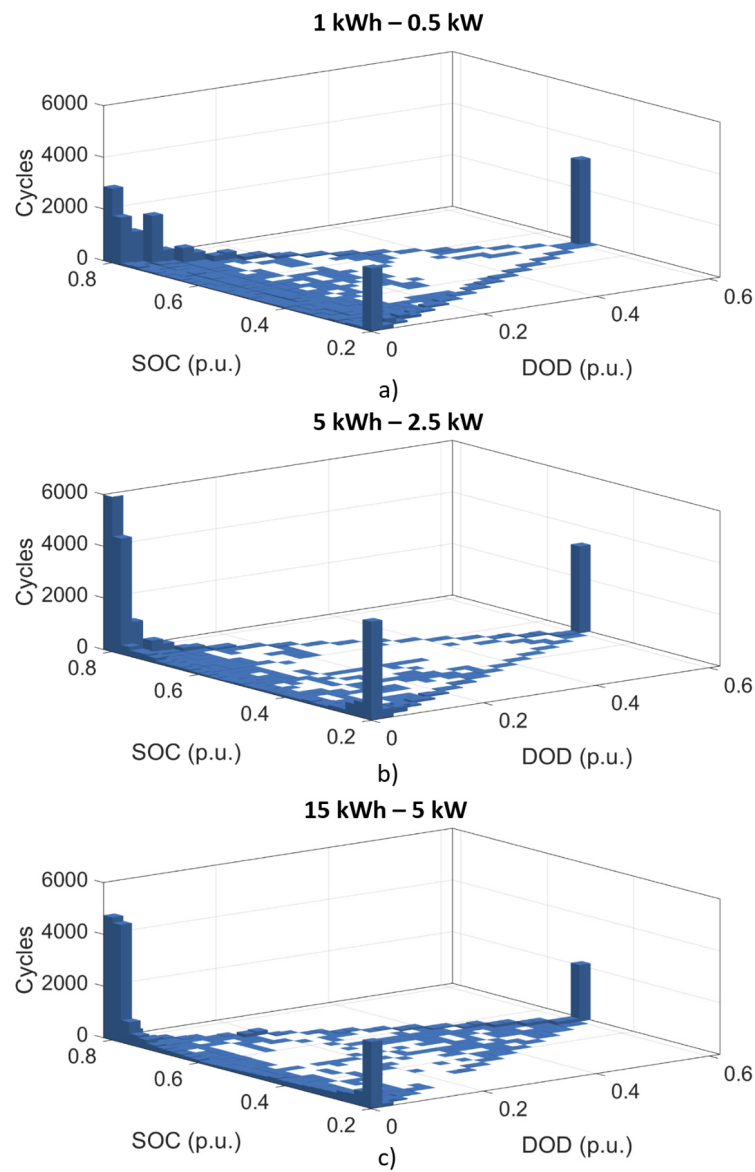

**Figure 5.** Cycles performed according to their depth of discharge and state of charge for different battery sizes. Source: self-elaborated.

This variation is attributed to differences in the charging capacity of the batteries and discharge energy, which impacts battery lifespan. A greater DOD per cycle accelerates battery degradation. Consequently, smaller capacity batteries, which complete higher

DOD cycles, experience more rapid aging and reduced lifespan compared to larger capacity batteries.

The economic analysis results for the different battery models are presented in Figure 6. The net present value is positive only for batteries with capacities of 5 kWh or less, with the highest value of 181.40 € achieved by the 2 kWh battery, making it the most profitable model. Regarding the DPB, it ranges from approximately 6 to 10 years, with the 2 kWh BESS yielding a DPB of 7.4 years. The DPB trend exhibits a steep slope relative to battery size, averaging 0.97 years/kWh. This indicates that selecting a battery with a smaller capacity significantly enhances the payback period.

The economic analysis conducted in [18] for homes with an annual consumption ranging from 5.9 to 9.6 MWh and an EOL of 80% resulted in a negative NPV for all battery sizes, indicating that such systems were unprofitable. In contrast, the proposed HEMS enhances the return on investment in BESSs. This improvement is attributed to the optimized energy dispatch and the use of a RTP tariff rather than a fixed tariff, which allows for effective energy arbitrage.

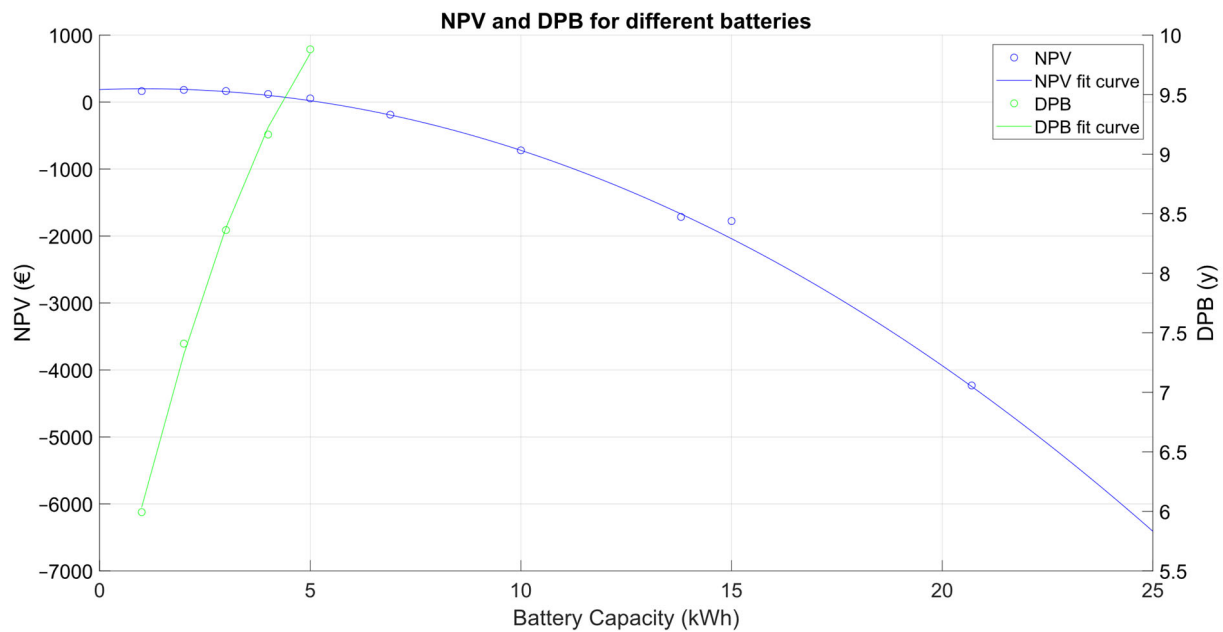

**Figure 6.** Economic analysis of battery sizing: NPV and DPB. Source: self-elaborated.

*3.2. Sensitivity Analysis of Time Resolution*

The proposed optimization method was applied over the lifetime of the BESS, as in the previous case, but in this scenario, using various time resolutions: 5, 15, 30, and 60 min. When simulating the operation of the BESS for time resolutions other than 5 min, the data is aggregated based on its average value over periods equal to the selected time resolution. The results of the technical study (including SCR, SSR and $T_{EOL}$) for different time resolutions are illustrated in Figure 7. The findings indicate that time resolution does not significantly affect the estimation of SCR and SSR. The variation in these ratios is minimal, with a maximum difference of 1.53% for SCR and 2.02% for SSR.

In contrast to the SCR and SSR, battery lifetime is significantly affected by the chosen temporal resolution. The higher the temporal resolution, the shorter the estimated battery lifetime. The difference between the results obtained for 5 and 60 min resolutions ranges from 0.68 to 1.77 years, resulting in an overestimation of lifetime by up to 20.61%.

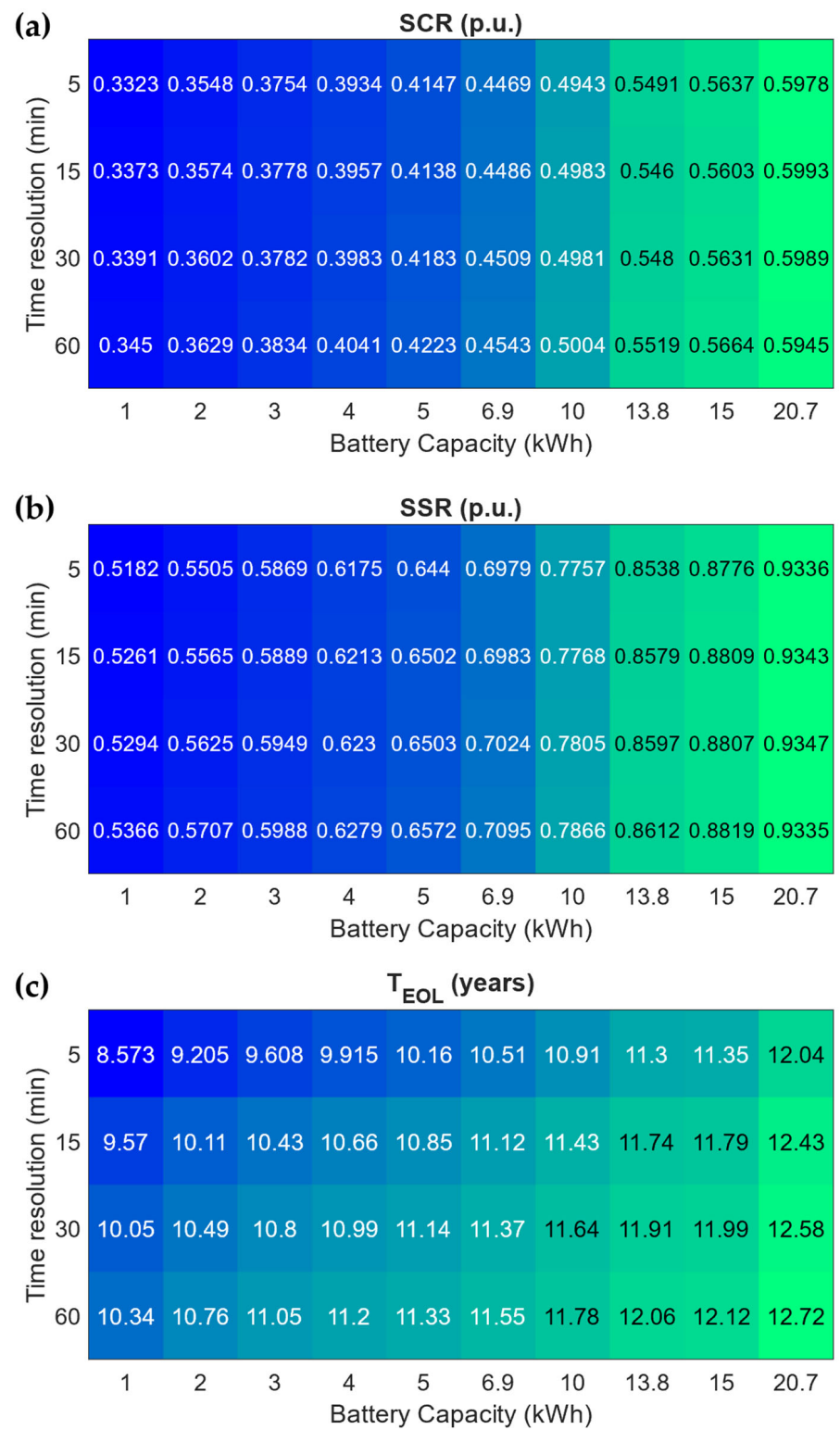

**Figure 7.** Sensitivity analysis of time resolution on technical parameters: (**a**) SCR, (**b**) SSR and (**c**) T$_{EOL}$. Source: self-elaborated.

In order to identify the reason why lifetime varies so much depending on time resolution, the RFC results have been analyzed. Figure 8 shows the cycles performed at a time resolution of 60 min, where the number of cycles performed at a DOD of 60% is prevalent.

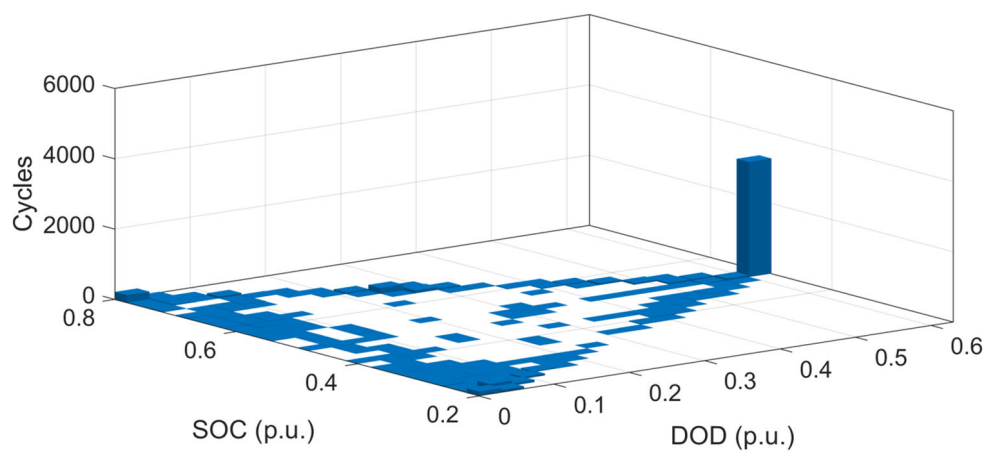

**Figure 8.** Cycles performed based on their depth of discharge and state of charge for a time resolution of 60 min. Source: self-elaborated.

Conversely, Figure 9, obtained with a resolution of 5 min, shows that the large majority of cycles are performed at a DOD of less than 10%. Although these cycles have a lower impact on battery lifespan compared to higher DOD cycles, the underestimation of their number is very significant when using a low time resolution. This leads to an underestimation of battery degradation, which translates into a substantial overestimation of the lifetime of the battery, as shown in Figure 7c.

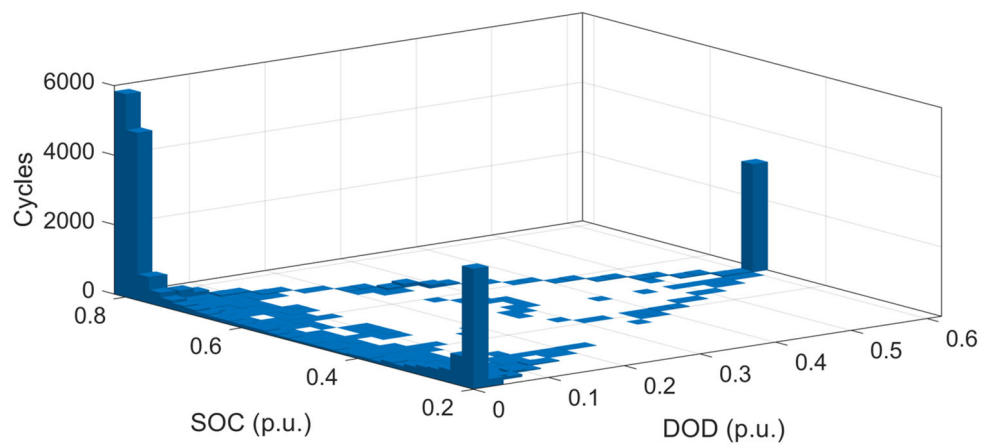

**Figure 9.** Cycles performed based on their depth of discharge and state of charge for a time resolution of 5 min. Source: self-elaborated.

The results of the economic study for different time resolutions are presented in Figure 10. As shown in Figure 10a, NPV varies significantly depending on the time resolution chosen, with higher resolutions resulting in lower NPV. The absolute decrease in profitability ranges from 289 € to 410 € for the 5 min resolution compared to the 60 min resolution, with relative decreases ranging from 7.33% to 184.68%.

A particularly notable case is the 6.9 kWh battery, which is profitable at temporal resolutions of 30 and 60 min but not at higher resolutions. This discrepancy arises from the underestimation of degradation at lower temporal resolutions, which leads to an overestimation of battery lifetime (as observed in Figure 7c). This finding highlights the susceptibility of economic feasibility studies to the effects of time resolution in energy management models that involve battery degradation.

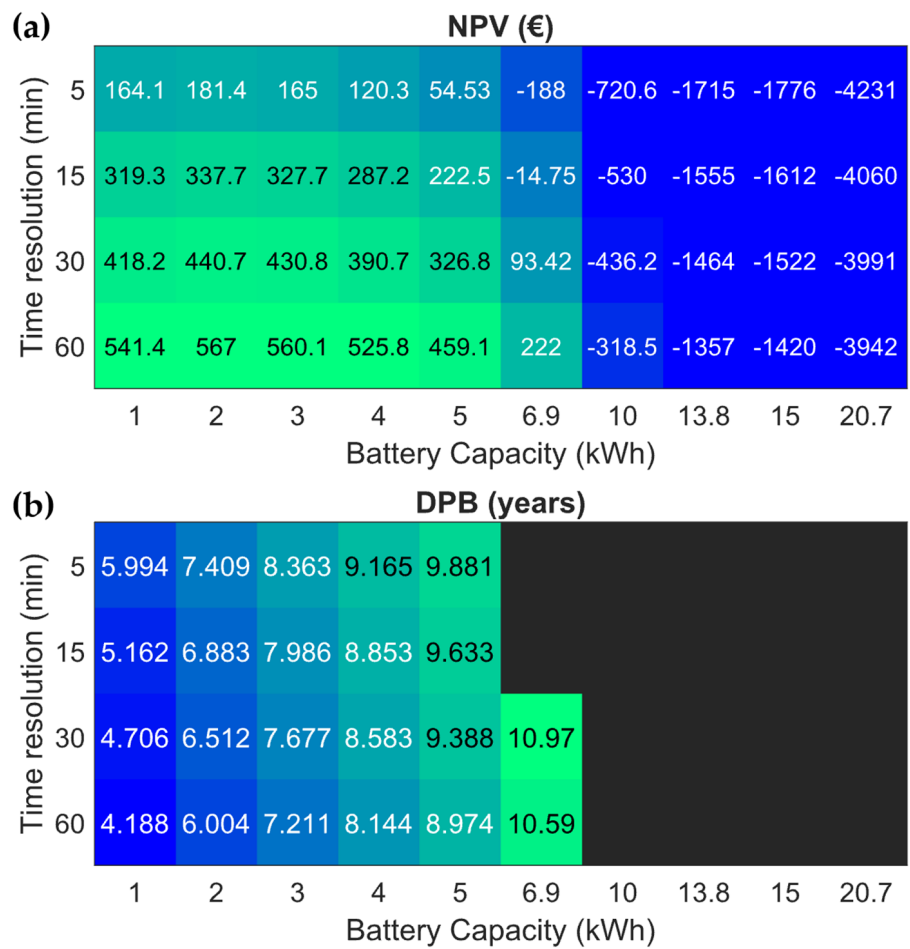

**Figure 10.** Sensitivity analysis of time resolution on economic parameters: (**a**) NPV and (**b**) DPB. Source: self-elaborated.

DPB, shown in Figure 10b, is also significantly affected by variation in temporal resolution. For a 5 min time resolution, the payback period increases by one to two years compared to the results obtained with a 60 min resolution. This increase in DPB makes the investment less attractive, as it extends the time required for the investment to break even, thereby potentially discouraging investors.

*3.3. Comparison of Current Operation with the Proposed Method*

Regarding the energy management method, the results for the 10 kWh battery in real operation mode compared to the proposed method showed significant differences in key metrics. As shown in Figure 11, the real operation of the battery achieved an SCR of 61.58% and an SSR of 94.17%. In contrast, the proposed optimization method achieved an SCR of 49.43% and an SSR of 77.57%. This notable difference arises primarily because the real operation mode was oriented towards maximizing self-consumption, which entails utilizing as much of the locally generated solar energy as possible. In contrast, the proposed method was designed to maximize the profitability and lifetime of the BESS. This approach may involve strategic charging and discharging to minimize the cost of energy imports and extend the lifespan of battery.

Furthermore, accounting for the cost of battery degradation results in a reduction in the energy delivered by the battery, which in turn decreases self-sufficiency. When compared to the SCR obtained in [18] for dwellings with similar consumption ranging from 30% to 65%, the values obtained in this study fall within this range.

The proposed power management method, as illustrated in Figure 11, extends the lifetime of batteries from 8.9 to 11.9 years, representing a 22.47% increase over the actual battery operation. This indicates that the proposed method effectively prolongs the lifespan of battery. The lifetime reported in [18] for BESSs ranges between 9 and 11 years, considering an EOL capacity of 80%, which aligns closely with the results obtained in this study.

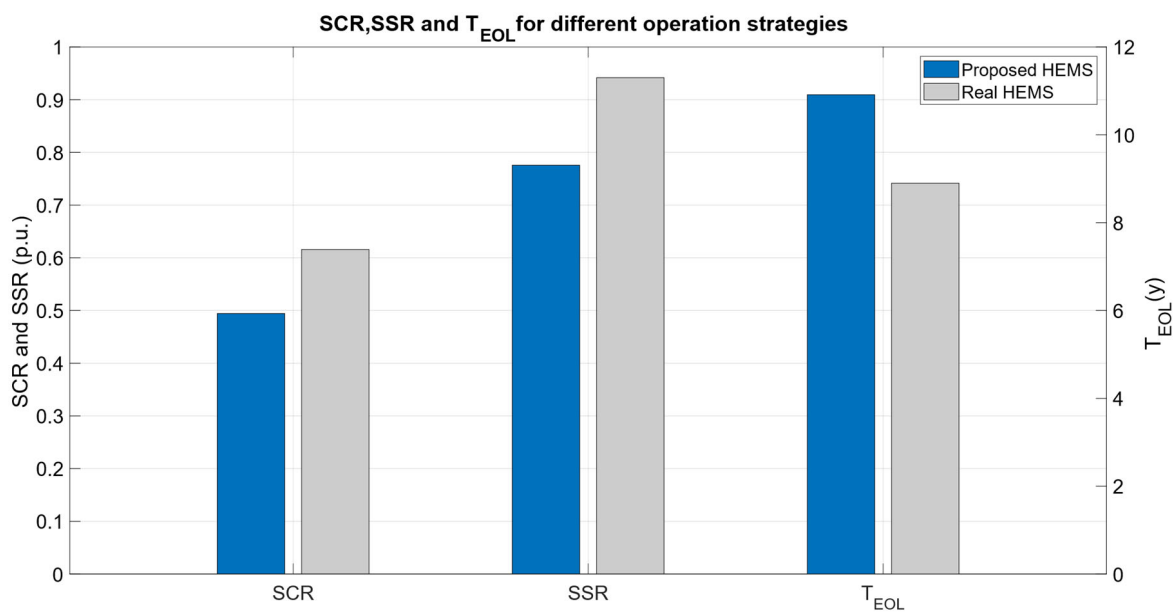

**Figure 11.** Technical analysis of real and proposed energy management for a 10 kWh BESS. Source: self-elaborated.

From an economic perspective, Figure 12 compares the proposed HEMS with the one currently implemented in the real system. The results indicate that the NPV is 21.29% higher than that of the real system (−720.60 € vs. −915.50 €). Although both NPVs are negative and thus not profitable, the proposed HEMS demonstrates an improvement in the investment's profitability.

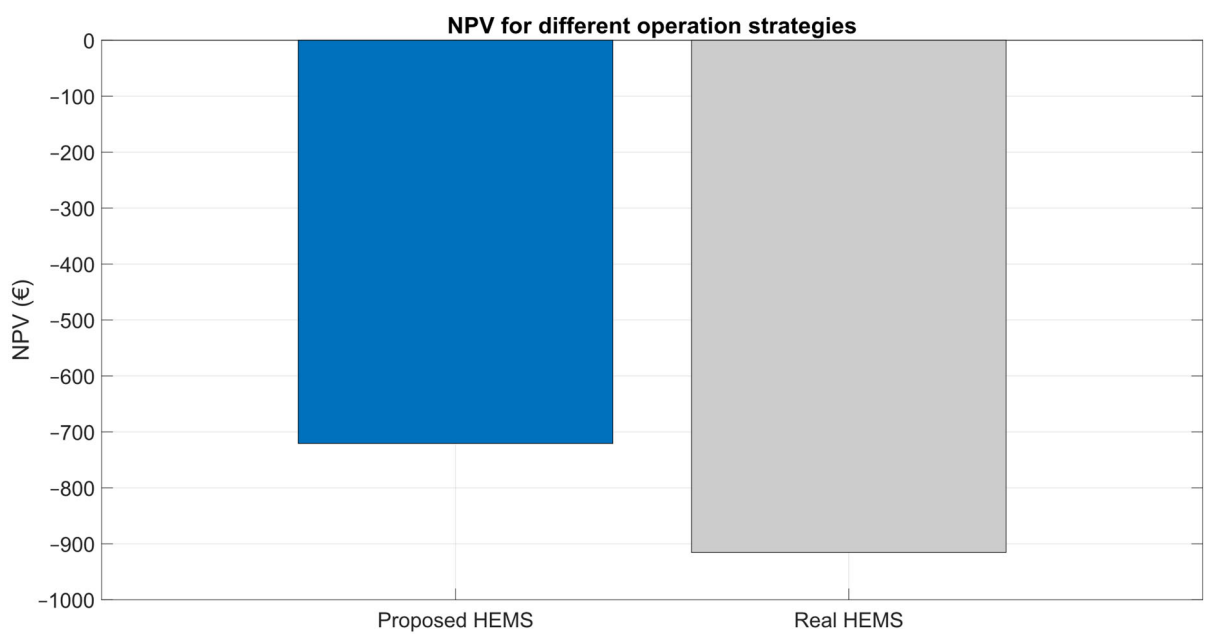

**Figure 12.** Economic analysis of real and proposed HEMSs for a 10 kWh BESS. Source: self-elaborated.

Future research will be conducted to validate the proposed HEMS algorithm. For this validation, programmable power electronics will be required, and it will need to be implemented in conjunction with a battery that has characteristics similar to those outlined in this study.

## 4. Conclusions

RTPVs encounter technical challenges, primarily due to the intermittent nature of solar power generation, which can compromise system stability and power quality. The variability in solar energy production results in fluctuations in power flow, significant challenges for power system management and integration. Moreover, RTPVs are confronted with economic challenges related to the depreciation of surplus solar energy. In numerous regions, the overproduction of PV added into the grid during peak irradiation hours surpasses local demand, resulting in a substantial decrease in electricity prices during these periods. To mitigate these issues, BESSs are proposed as a viable solution. BESSs can enhance the efficiency and cost-effectiveness of RTPVs, improving both their viability and sustainability.

Within the residential sector, this study focuses on both homes in which a RTPV is incorporated or those with a previous photovoltaic installation and the potential for DC coupling with a battery. For the latter, upgrading the configuration of these installations results in improved self-consumption and self-sufficiency, thereby increasing the utilization of the photovoltaic installation. To evaluate this methodology, it was applied to a real case, and the results were compared across various energy management strategies. This analysis sought to determine the optimal size of the BESS and assess how different time resolutions impact the results.

The BESS sizing was determined through a tecno-economic analysis. The NPV values for profitable BESS sizes (ranging from 1 to 5 kWh) varied from 54.53 € to 181.40 €, indicating modest financial returns. The DPB varied between 6 and 10 years, with shorter payback periods associated with smaller battery sizes. These results suggest that BESSs can serve as a cost-effective solution for enhancing PV installations, with the real case study achieving a self-consumption rate of up to 41.47% and a self-sufficiency rate of 64.40%.

With the proposed HEMS, the lifespan of the current BESS increased by up to 22.47% compared to the current operational approach. This extension in battery life enhanced the cost-effectiveness of the BESS by 21.29%. Therefore, the proposed HEMS not only improves the profitability of the BESS but also prolongs its operational lifespan relative to the methods currently available in the scientific literature on battery energy management.

The sensitivity analysis of temporal resolution revealed a substantial impact on the economic evaluation of the BESS. Specifically, using a 5 min resolution compared to a 60 min resolution results in a reduction of up to 184.68% in the NPV and an increase of up to 43.12% in the DPB for cost-effective BESSs. This highlights the importance of temporal resolution in BESS sizing studies, especially when the energy management method accounts for battery. This approach, which balances the computational demands of degradation estimation with the optimization of the objective function, represents a novel contribution to the field. It contrasts with previous studies where temporal resolution either was not considered or was set at excessively high values.

The study demonstrates that investing in batteries is economically viable, especially when supported by advanced energy management methods, such as the one proposed. However, the profitability remains modest and is sensitive to uncertainties in factors like discount rate and energy costs. Thus, incentives and cost reductions are essential to ensure the economic feasibility of battery investments. This study highlights the importance of commercially available battery sizes of 5 kWh or less, which could enhance the financial attractiveness of battery-integrated rooftop PV systems. Offering smaller capacity batteries could optimize BESS size, increase return on investment, and drive higher adoption rates for new residential installations while benefiting upgrades to existing systems.

**Author Contributions:** All authors contributed equally to this work. All authors have read and agreed to the published version of the manuscript.

**Funding:** The authors would like to thank the program which finances projects aimed at ecological and digital transition (Grant No. TED2021-131137B-IO0: "Contribution to the Ecological Transition of the Industrial Sector through Photovoltaic Self-consumption") and the research program "Investigo" (2022-c23-i01-p03.s0020-0000590) which has been financed by the European Union with Next Generation EU funds. The authors also acknowledge the support provided by the Thematic Network 723RT0150 "Red para la integración a gran escala de energías renovables en sistemas eléctricos (RIBIERSE-CYTED)" financed by the call for Thematic Networks of the CYTED (Ibero-American Program of Science and Technology for Development) for 2022.

**Data Availability Statement:** The data has been obtained from a private home with the consent of the owner but is not available for security reasons.

**Acknowledgments** The authors would like to thank the Andalusian company EFITRON, for providing technical information and installation data used in this study.

**Conflicts of Interest:** The authors declare no conflicts of interest.

## Abbreviations

| | |
|---|---|
| AC | Alternating current |
| BESS | Battery energy storage system |
| BTM | Behind the meter |
| DC | Direct current |
| DCF | Discounted cash flow |
| DG | Distributed generation |
| DOD | Depth of discharge |
| DPB | Discounted payback period |
| EOL | End of Life |
| FIT | Feed-in tariff |
| FTM | Front of the meter |
| HEMS | Home energy management system |
| IRR | Internal rate of return |
| LCOE | Levelized cost of energy |
| LFP | Lithium iron phosphate |
| MILP | Mixed integer linear programming |
| MPPT | Maximum power point tracker |
| NPV | Net present value |
| PV | Solar photovoltaic |
| RES | Renewable energy systems |
| ROI | Return on investment |
| RTP | Real-time price |
| RTPV | Rooftop photovoltaic system |
| SCR | Self-consumption ratio |
| SEI | Solid-electrolyte interface |
| SOC | State of charge |
| SOH | State of health |
| SSR | Self-sufficiency ratio |
| TOU | Time-of-use |
| VAT | Value-added tax |
| VPSC | Voluntary price for small customer |

## Appendix A

In this section, the applied degradation model of the battery (extracted from [40]) is described. The capacity loss $L_{k_{op}}$ depends on the degradation factor $f_{b,k_{op}}$ and the solid-electrolyte interface (SEI) parameters as indicated in Equation (A1). The degradation factor was determined by combining the degradation due to battery cycling and aging as described in Equation (A2) from the beginning of the simulation to the current optimization period (A3).

$$L_{k_{op}} = 1 - \alpha_{sei}\ e^{-f_{b,k_{op}}\beta_{sei}} - (1 - \alpha_{sei})\ e^{-f_{b,k_{op}}} \tag{A1}$$

$$f_{b,k_{op}} = \sum_{r_{op}} f_{c,r_{op}}\ c_{r_{op}} + f_{t,r_{op}} t_{r_{op}} \tag{A2}$$

$$r_{op} = t \in (0, k_{op}T_{op}) \tag{A3}$$

Each degradation factor was calculated based on various variables affecting battery degradation as shown in (A4) and (A5), specifically the battery temperature $T_b$, the number of cycles performed $c$, their depth of discharge $\delta$, the average SOC $\sigma$, and the time elapsed since the battery was put into operation $t$. The stress factors were calculated as a function of these variables by Equations (A6)–(A9) and coefficients obtained through experimental data: $k_T$, $k_{\delta 1}$, $k_{\delta 2}$, $k_{\delta 3}$, $k_\sigma$, $k_t$.

$$f_{c,r_{op}} = S^T(T_{b,r_{op}})\ S^\delta(\delta_{r_{op}})\ S^\sigma(\sigma_{r_{op}}) \tag{A4}$$

$$f_{t,r_{op}} = S^T(T_{b,r_{op}})\ S^t(t_{r_{op}})\ S^\sigma(\sigma_{r_{op}}) \tag{A5}$$

$$S^T(T_{b,r_{op}}) = e^{k_T(T_{b,r_{op}} - T_{b,r})\frac{T_{b,r}}{T_{b,r_{op}}}} \tag{A6}$$

$$S^\delta(\delta_{r_{op}}) = \left(k_{\delta 1}\delta_{r_{op}}^{k_{\delta 2}} + k_{\delta 3}\right)^{-1} \tag{A7}$$

$$S^\sigma(\sigma_{r_{op}}) = e^{k_\sigma(\sigma_{r_{op}} - \sigma_r)} \tag{A8}$$

$$S^t(t_{r_{op}}) = k_t t_{r_{op}} \tag{A9}$$

The number of cycles performed, their DOD, and the average SOC were obtained by applying the RFC algorithm to the state of charge within the $r_{op}$ period. The reference battery temperature $T_{b,r}$ was considered to be 25 °C and the reference state of charge $\sigma_r$ was set at 50%. Table A1 presents the values used for the battery degradation model, as extracted from [41].

**Table A1.** Battery degradation model parameters. Source: self-elaborated.

| Symbol | Value | Units |
|---|---|---|
| $\alpha_{sei}$ | $5.75 \times 10^{-2}$ | - |
| $\beta_{sei}$ | 121 | - |
| $k_T$ | $6.93 \times 10^{-2}$ | - |
| $k_{\delta 1}$ | $1.40 \times 10^5$ | - |
| $k_{\delta 2}$ | $-5.01 \times 10^{-1}$ | - |
| $k_{\delta 3}$ | $-1.23 \times 10^5$ | - |
| $k_\sigma$ | 1.04 | - |
| $k_t$ | $4.14 \times 10^{-10}$ | $s^{-1}$ |
| $T_{b,r}$ | 25 | °C |
| $\sigma_r$ | 0.5 | p. u. |

Actividades de Transporte y Distribución de Energía Eléctrica; Boletín Oficial del Estado: 2019; Vol. 279–I, pp 127725–127734. Available online: https://www.cnmc.es/sites/default/files/2749227_42.pdf (accessed on 20 May 2024).